\documentclass[preprint,12pt]{elsarticle}
\usepackage{graphicx}
\usepackage{amsmath}
\usepackage{amssymb}
\usepackage{amsthm}
\usepackage{booktabs}
\usepackage{stmaryrd}
\usepackage{longtable}
\usepackage[figuresright]{rotating}
\usepackage{multirow}
\usepackage{dcolumn}
\usepackage{anysize}
\usepackage{color}
\marginsize{1.5cm}{1.5cm}{1.5cm}{1.5cm}
\newcolumntype{d}[1]{D{.}{.}{#1}}
\begin{document}
\title{Crystal field of rare earth impurities in LaF$_3$}
\author{P. Nov\'ak\corref{cor1}}
\cortext[cor1]{Corresponding author}
\ead{novakp@fzu.cz}
\author{J. Kune\v{s},  K. Kn\'i\v{z}ek} 
\address{Institute of Physics of ASCR, Cukrovarnick\'a 10, 162 00 Prague 6, Czech Republic}
\date{\today}
\begin{abstract}
The crystal field parameters of 13 trivalent lanthanide ions substituted for La in  LaF$_3$
were calculated using the combination of the band structure and Wannier function calculations.
Performing an atomic exact diagonalization with thus obtained crystal-field parameters we
compute the crystal-field splitting of atomic multiplets.
The calculation is compared with the available experimental results and a good agreement is found.
\end{abstract}
\maketitle
\section{Introduction}
Lanthanum trifluoride doped with the rare-earth (R) ions is commonly used for the optical applications and
as a material for electrodes in detectors. During the past decades the optical spectra of all R impurities except Pm in
LaF$_3$ were measured and the results analyzed using the semiempirical crystal field models \cite{carnall1,carnall2}. 
 With the  exceptions of the work 
of Ishii {\it et al.} \cite{ishii} and
Brik {\it et al.} \cite{brik}, based on the
discrete variational multi-electron method (DV-ME), the {\it ab-initio} theory was missing, however.
DV-ME method and its results are discussed and compared to
the present ones in part 5. The survey of the attempts to calculate
the crystal-field parameters (CFP) in the rare-earth compounds may be found in 
\cite{ogasawara,novak4}. 

A new theoretical approach has been recently used to calculate the crystal field parameters  of
rare-earth in aluminates, gallates, cobaltites and manganites with the orthorhombic perovskite structure
 \cite{novak1,novak2,novak3}. The method uses
the density functional theory  band structure calculation, followed by a transformation of the
Kohn-Sham Hamiltonian to the Wannier basis and an expansion of its local part in the spherical tensor
operators. The CFP serve as an input to an atomic exact diagonalization program, which 
takes into account also the $4f$-$4f$ interaction, spin-orbit coupling and the Zeeman
interaction.  The calculated results agree  remarkably well with the experiment: the 
crystal-field-split 
multiplet levels within a few meV and magnetic properties are correctly described as well.  
For any site symmetry the method yields an unambiguous set of CFP.
 This is important especially if the local symmetry of the R site is low and many CFP are needed to
fully characterize the crystal field. Semiempirical 
methods, as well as the analysis of optical data make usually use of the
 least squares fitting, a procedure often yielding ambiguous results.
With the exception of a single parameter, to be discussed below, the present method is fully {\it ab-initio} i.e. the only
necessary inputs are the atomic composition and the crystal structure of the compound in question. 

In the orthorhombic perovskites, to which the method was already successfully applied  \cite{novak1,novak2,novak3},
 the symmetry of the R site is $C_s$ and 15 parameters, 3 real and 6 complex CFP,
are necessary. The situation of R impurities
in  LaF$_3$, which is considered here,   is similar. The site symmetry of
La$^{3+}$ site in LaF$_3$ is $C_2$ and again 15 parameters, 3 real and 6 complex CFP, are necessary.

The paper is organized as follows: the method and the computational details 
are described in sections 2 and 3, respectively. 
The calculated energy levels and CFP are presented in section 4. In the same 
section the theoretical crystal 
field spectra are compared with the experimental data collected by Carnall {\it et al.} \cite{carnall1,carnall2}. 
Section 5 contains discussion, followed by the conclusions in section 6.

\section{Methods}
\label{sec:methods}
An effective Hamiltonian operating on the $4f$ electrons can be written as
\begin{equation}
\label{eq:h}
\hat{H}_{eff}=\hat{H}_A + \hat{H}_Z + \hat{H}_{CF},
\end{equation}
where $\hat{H}_A$ is the spherically symmetric free ion Hamiltonian, $\hat{H}_Z$ corresponds to the Zeeman
interaction and $\hat{H}_{CF}$ is the crystal-field term. In the Wybourne notation \cite{wybourne} $\hat{H}_{CF}$
has the form
\begin{equation}
\hat{H}_{CF} = \sum_{k=0}^{k_{max}}\sum_{q=-k}^k B_{q}^{(k)} \hat{C}_{q}^{(k)}, \;\;
 \hat{C}_{q}^{(k)}= \sum_{i,j=1}^{7} [C_{q}^{(k)}]_{ij} f_{i}^{\dagger} f_{j}
\label{eq:hcf}
\end{equation}
where $ \hat{C}_{q}^{(k)}$ is a spherical tensor operator of rank $k$ acting on the
$4f$ electrons of the R ion, for which $k_{max}$ is equal to six. The coefficients $B_{q}^{(k)}$ are
the CFP. Hermiticity of $\hat{H}_{CF}$ requires that $(B_{-q}^{k})^* = (-1)^q B_{q}^{k}$.
The operators $f_{i}^{\dagger},\, (f_{i})$ create and annihilate an electron in an 
$f$-orbital $i$.
The details of $\hat{H}_A$ are given e.g. in Ref. \cite{hufner}.

Calculation of the CFP proceeds in four steps:
\begin{enumerate}
 \item {Standard selfconsistent band calculation with $4f$ states included in the core. The result
of this step is the Kohn-Sham potential, which is subsequently used in the next step. The calculation
is non spin-polarized.} 
\item{ A Hamiltonian describing R $4f$ states together with fluorine $2p$ and $2s$ states 
subject to the Kohn-Sham potential from the previous step is diagonalized. 
To eliminate other valence states an orbital operator
which adds a large positive constant to the potential acting on these states is used.
Relative position of $4f$ and fluorine states is adjusted by applying similar operator to the
fluorine states, which shifts their potential by the 'hybridization' parameter $-\Delta$
(the only parameter of method).} 
\item { The $4f$ band states are transformed to Wannier basis using the
wien2wannier \cite{w2w} and wannier90 \cite{wannier} packages. 
 The result of this step, relevant for 
CFP calculation is the Hamiltonian $\hat{H}_W$ in the basis of $4f$ Wannier functions.}
\item{Local $4f$ Hamiltonian
$\hat{h}^{loc}$ is obtained as the on-site part $\hat{H}_W$. 
 Finally $\hat{h}^{loc}$ is expanded in series of
spherical tensor operators. The coefficients of expansion are the CFP $B^{(k)}_q$.}
\end{enumerate}
To find $B^{(k)}_q$ we made use of the fact that $\hat{C}^{(k)}_q$ form complete orthogonal 
set of operators in the
subspace of $4f$ states. Then
\begin{equation}
\label{eq:bcalc}
 B^{(k)}_q = \frac{1}{n_{k,q}} \sum_{i=1}^7 \sum_{j=i}^7 h^{loc}_{ij} [\bar{C}^{(k)}_{q}]_{(ij)} ,
\end{equation}
where $n_{k,q}$ is the normalizing factor:
\begin{equation}
 n_{k,q} =\sum_{i=1}^7 \sum_{j=i}^7 [\bar{C}^{(k)}_{q}]_{(ij)} [C^{(k)}_{q}]_{(ij)} .
\end{equation}

With the CFP in hand we use a modified 'lanthanide' package \cite{lanthanide}
to solve the eigenvalue problem for Hamiltonian (1) and  calculate the splitting of multiplets by the crystal field.
Detailed description of the analysis may be found in \cite{novak1,novak2}.

\subsection{Hybridization parameter}
\label{sec:delta}
The parameter $\Delta$ appears due to the hybridization between the rare-earth $4f$ states and the
valence states of its ligands. In the $3d$ metal compounds the hybridization is in most cases a dominating source of the
crystal field. For the Co:ZnO system Kuzian {\it et al.} \cite{kuzian} successfully reproduced the $g$-factors and the
zero-field splitting, by treating the hopping between the cobalt $3d$ orbitals and the $2p$ states of the fluorine ligands as a 
perturbation. 
Our treatment of the hybridization is analogical to that of Kuzian {\it et al.} and is briefly described in Ref. \cite{novak1}.
$\Delta$ can be estimated using a charge transfer energy
\begin{equation}
\label{eq:delta}
 \Delta \simeq E_{tot}(4f^{(n+1)},N_{val} - 1) - E_{tot}(4f^n,N_{val}),
\end{equation}
where $E_{tot}(4f^n,N_{val})$ is the total energy of the ground state of the system ($n_{4f}$ electrons in $4f$ shell of R ion and
$N_{val}$ electrons in the valence band), while $E_{tot}(4f^{(n+1)},N_{val} - 1)$ corresponds to the excited state in which
one of the valence electrons was transferred in the $4f$ electron shell. The hybridization parameter thus can be calculated
by performing two calculations with $4f$ electrons treated as the core states - the first one with $ 4f^n,N_{val}$, the second 
with $ 4f^{(n+1)},N_{val} - 1$ electron configurations. The results of such a calculation for R:LaF$_3$ system are presented in
section \ref{sec:results}.

\section{Details of calculation}
\label{sec:calc}
\begin{figure}
\begin{center}
\includegraphics[width=12cm]{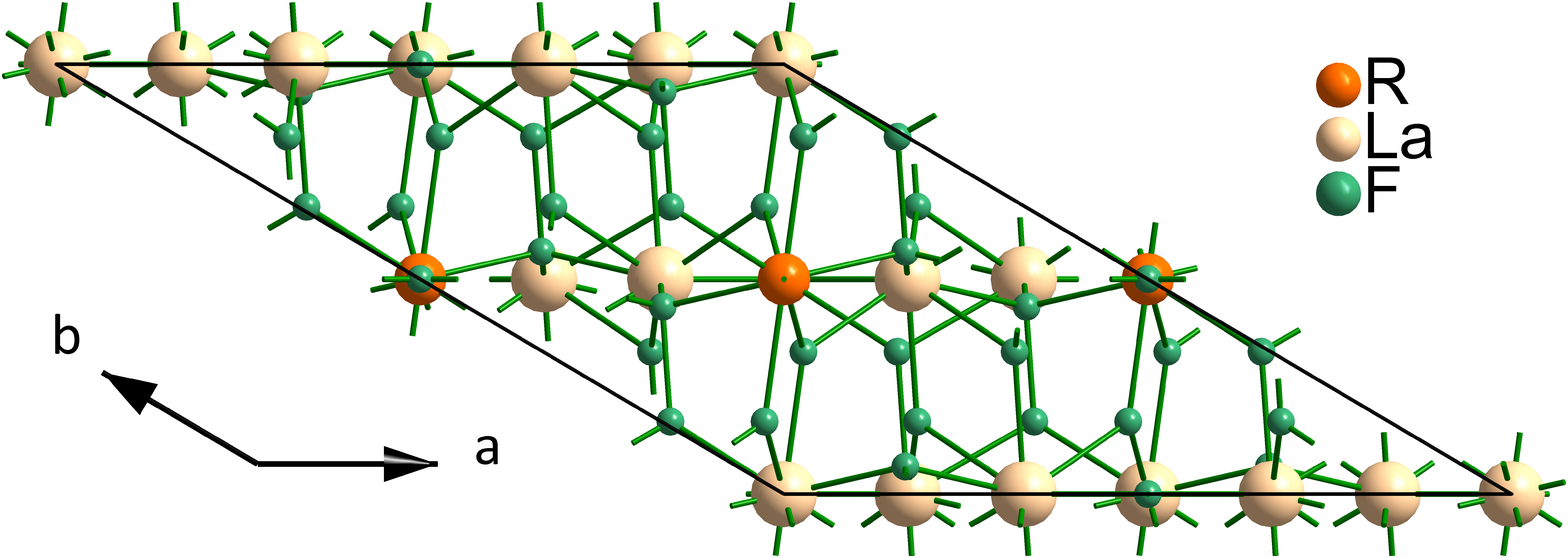}
\caption{(Color online)Unit cell of R$_{1/6}$La$_{5/6}$F$_3$ used in the calculation of CFP. The $C_2$ symmetry axis
is along the axis a.}
\label{fig:structure}
\end{center}
\end{figure}

The LaF$_3$ crystallizes in trigonal symmetry ($P\bar{3}c1$ space group) with six La in the       
hexagonal unit cell \cite{zalkin}.
To determine the band structure (steps 1 and 2 in the previous section) the WIEN2k package \cite{wien} was used
with the exchange-correlation of the generalized-gradient approximation form \cite{perdew}. 
For the pure LaF$_3$ compound the experimental crystal structure was taken from \cite{zalkin}. In R:LaF$_3$
one of the six La atoms was replaced by a rare-earth atom and the internal structure was optimized by minimizing the atomic forces.
The resulting structure has only the identity and $C_2$ symmetry operations. The unit cell containing 24 
atoms is displayed in Fig. \ref{fig:structure}.
The eigenvalue problem was solved in 47 points of the irreducible part of the  Brillouin zone and 
the number of basis functions was $\sim$ 1780 (corresponding to parameter $RK_{max}$=7). The
calculations were non-spin-polarized. The atomic radii of R and F were 2.4 and 2 a.u., respectively.

The multiplet splitting and magnetism of rare-earth ions depend only weakly on the free ion parameters,
though there may be exceptions when the energy difference of the free ion multiplet is small.
Carnall {\it et al.} \cite{carnall1,carnall2} analyzed systematically the available absorption and 
luminescence data of trivalent lanthanide ions
doped into LaF$_3$. Using the least squares analysis the free ion parameters 
were obtained, which we used in  Ref. \cite{novak1,novak2,novak3}, as well as here. 
In addition 9 real CFP were obtained, corresponding to the $C_{2v}$ symmetry by which the actual $C_2$ symmetry was approximated.

 The comparison between experimentally observed and calculated crystal field splittings
of the  $|L,S,J,M_J\rangle$ multiplets deserves special attention. 
The experimental data reported by different authors are summarized in \cite{carnall2}. Understandably not all energy 
levels were detected (for Pm the experimental data are missing altogether). This makes the comparison of theory and experiment 
complicated as in many cases the missing levels are not unambiguously identified. Moreover, the crystal field
could lead to overlap of different $|L,S,J,M_J\rangle$ multiplets. 
Faced with this situation we characterize the agreement of theory and experiment by the mean square deviation
\begin{equation}
\label{eq:chi}
 \chi = \left [\frac{1}{N_{\mathrm{detected}}}  \sum_{i=1}^{N_{\mathrm{group}}} \sum_{j=1}^{n_i} (E_{i,j}^{exp} - E_{i,j}^{calc})^2 \right]^{1/2} ,
\end{equation}
where $N_{\mathrm{group}}$ equals to the number of multiplets with the provision that overlapping multiplets
are treated as a single group. $n_i$ is the number of states detected in the $i$th group and
\begin{equation}
 N_{\mathrm{detected}} = \sum_{i=1}^{N_{\mathrm{group}}} n_i .
\end{equation}
The energies $E_{i,j}^{exp}$ and $E_{i,j}^{calc}$ are the experimental and calculated energies, respectively, of the $j$th
eigenstate belonging to the $i$th group. These energies are taken relative to the lowest energy of the
group, which minimizes the influence of the free ion parameters. When some
of the eigenstates are not experimentally detected several ways to match  $E_{i,j}^{calc}$ and $E_{i,j}^{exp}$ are possible.
We did not attempt to minimize $\chi$ by probing all such pairings, rather the correspondence used by
Carnall {\it et al.} \cite{carnall2} was adopted. In Table I the numbers of experimental levels, which we 
below compare with those calculated, are given for each R, except Pm.

\begin{table}
\caption{R:LaF$_3$. $N_{\mathrm{detected}}$ is number of experimentally detected levels, energies of which 
are compared with the calculation. $N_{\mathrm{group}}$ is number of groups (isolated or overlapping multiplets)  
from which these levels originate. $N_{\mathrm{total}}$ is total number (detected and undetected) of levels in
$N_{\mathrm{group}}$ groups.}
\centering
\begin{tabular}{ccccccccccccc}
\hline
\hline
R & Ce  &  Pr   &  Nd  &  Sm  &  Eu  &   Gd &  Tb  &  Dy & Ho & Er & Tm & Yb \\
 \hline
$N_{\mathrm{group}}$    & 2 & 11 & 22 & 15  & 10 & 10  & 11 & 14 & 16  & 17 & 12 & 2 \\
$N_{\mathrm{detected}}$ & 7 & 75 & 104 & 65  & 31  & 54 & 51 & 77 & 137 & 80 & 56 & 6 \\
$N_{\mathrm{total}}$    & 7 & 90 & 115 & 68  & 59 & 57 & 99 & 84 & 170  & 84  & 90 & 7 \\
\hline
 \hline
\end{tabular}
\label{tab:n}
\end{table}

\section{Results}
\label{sec:results}
\subsection{Energy levels}
\begin{figure}
\includegraphics[width=12cm]{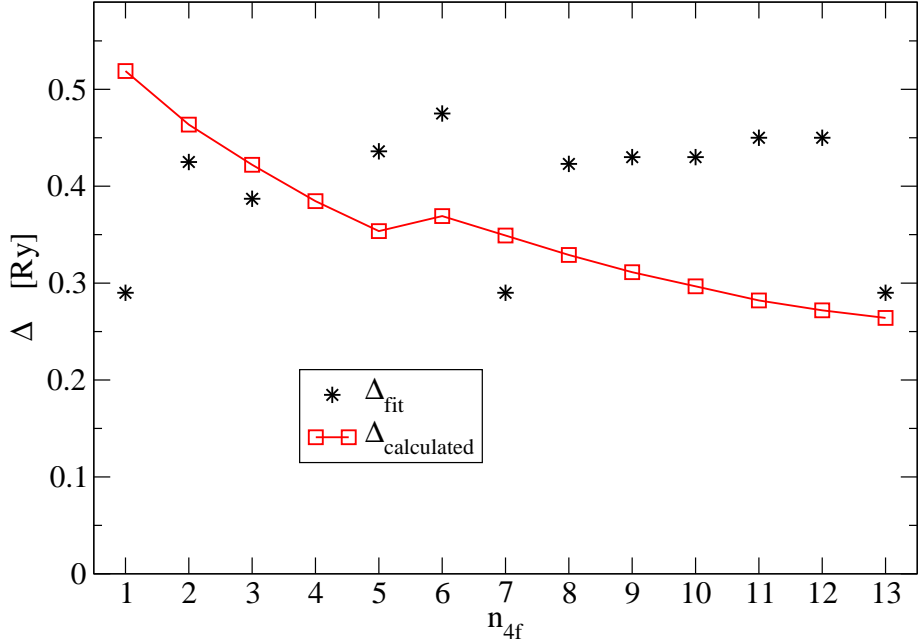}
\caption{(Color online)Comparison of the hybridization parameter $\Delta$ calculated using eq. \ref{eq:delta} with the
value which minimizes the mean square deviation $\chi$ (eq. \ref{eq:chi}). The curve in this, as well as in the following
figures,serves as a guide for the eyes only.}
\label{fig:delta}
\end{figure}

\begin{figure}
\includegraphics[width=12cm]{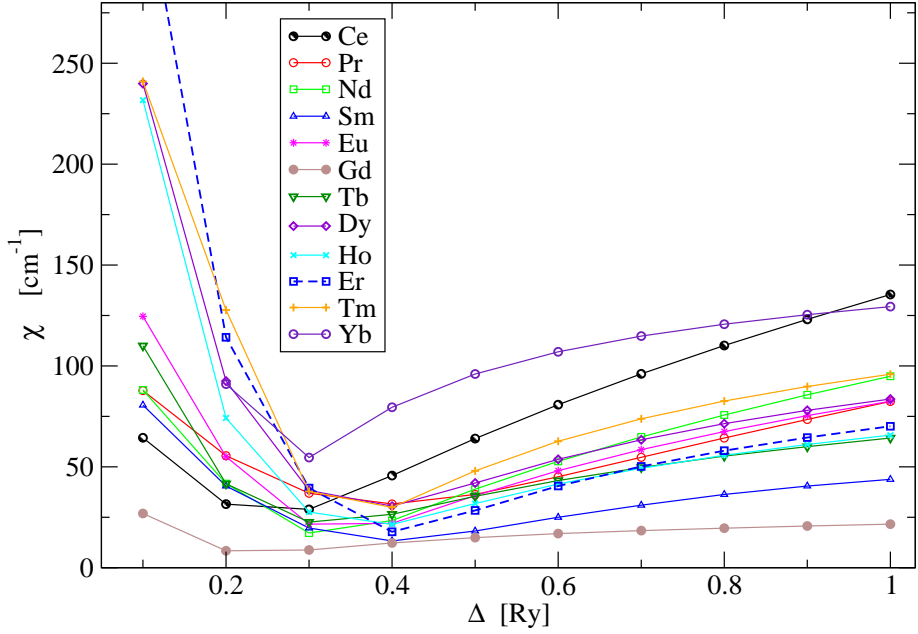}
\caption{(Color online) The mean square deviation $\chi$ (eq. \ref{eq:chi}) as function of the hybridization parameter $\Delta$.}
\label{fig:chi}
\end{figure}

\begin{figure}
\includegraphics[width=12cm]{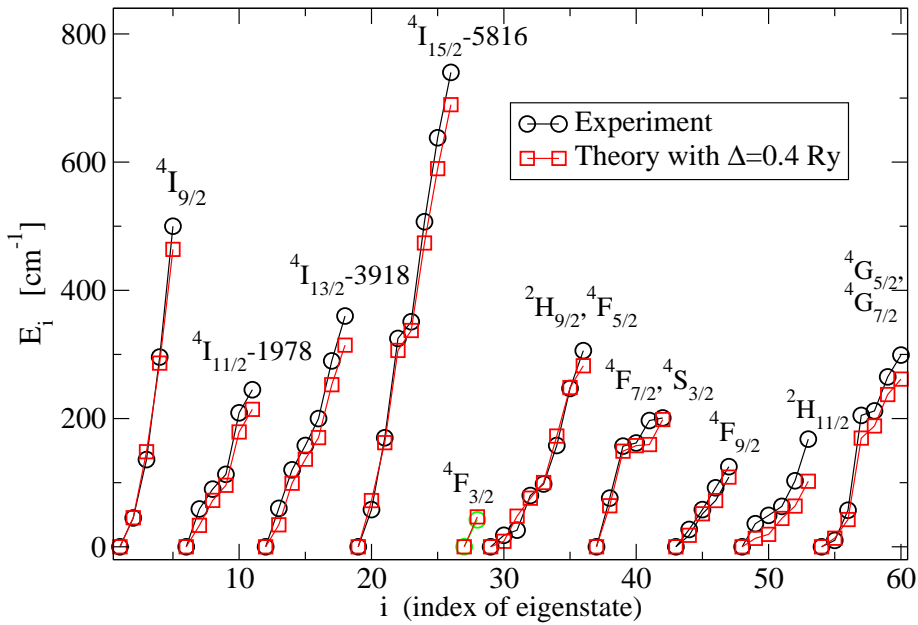}
\caption{(Color online) Nd:LaF$_3$. Comparison of experimental and calculated crystal field
splitting for $\Delta$ = 0.4 Ry and ten low lying multiplets. The energy, which was subtracted from the lowest eigenenergies
of the four lowest multiplets, is explicitly given in the figure (units of cm$^{-1}$), for six higher lying groups of
levels corresponding values are 11592, 12596, 13514, 14834 and 15997 cm$^{-1}$. Experimental data were taken from
Ref. \cite{carnall2}.}
\label{fig:ENd}
\end{figure}

The atomic program yields energies of all eigenstates arising from the electron configuration $4f^n$ 
(there are 7 such eigenstates for $n_{4f}$=1 and 13, maximum number 3432 is reached for $n_{4f}$=7, 
corresponding to Gd$^{3+}$ ion). We calculated the energies as a function of the hybridization parameter $\Delta$,
which was varied between 0.1 to 1 Ry with a step 0.1 Ry. By comparison with the experimental data we found 
that optimal $\Delta$ lies between 0.3 and 0.5 Ry. In Fig. \ref{fig:delta}, the optimal $\Delta$ is compared
to $\Delta$ calculated using (\ref{eq:delta}).
The dependence of the mean square deviation $\chi$ (\ref{eq:chi}) on the $\Delta$ is shown in Fig. \ref{fig:chi} for 
all R except Pm.  For Nd$^{3+}$ all of the levels corresponding to the  ten low lying 
multiplets were detected and as an example the
comparison of theory and experiment is shown in Fig. \ref{fig:ENd}.

\subsection{Crystal field parameters}
\begin{figure}
\includegraphics[width=12cm]{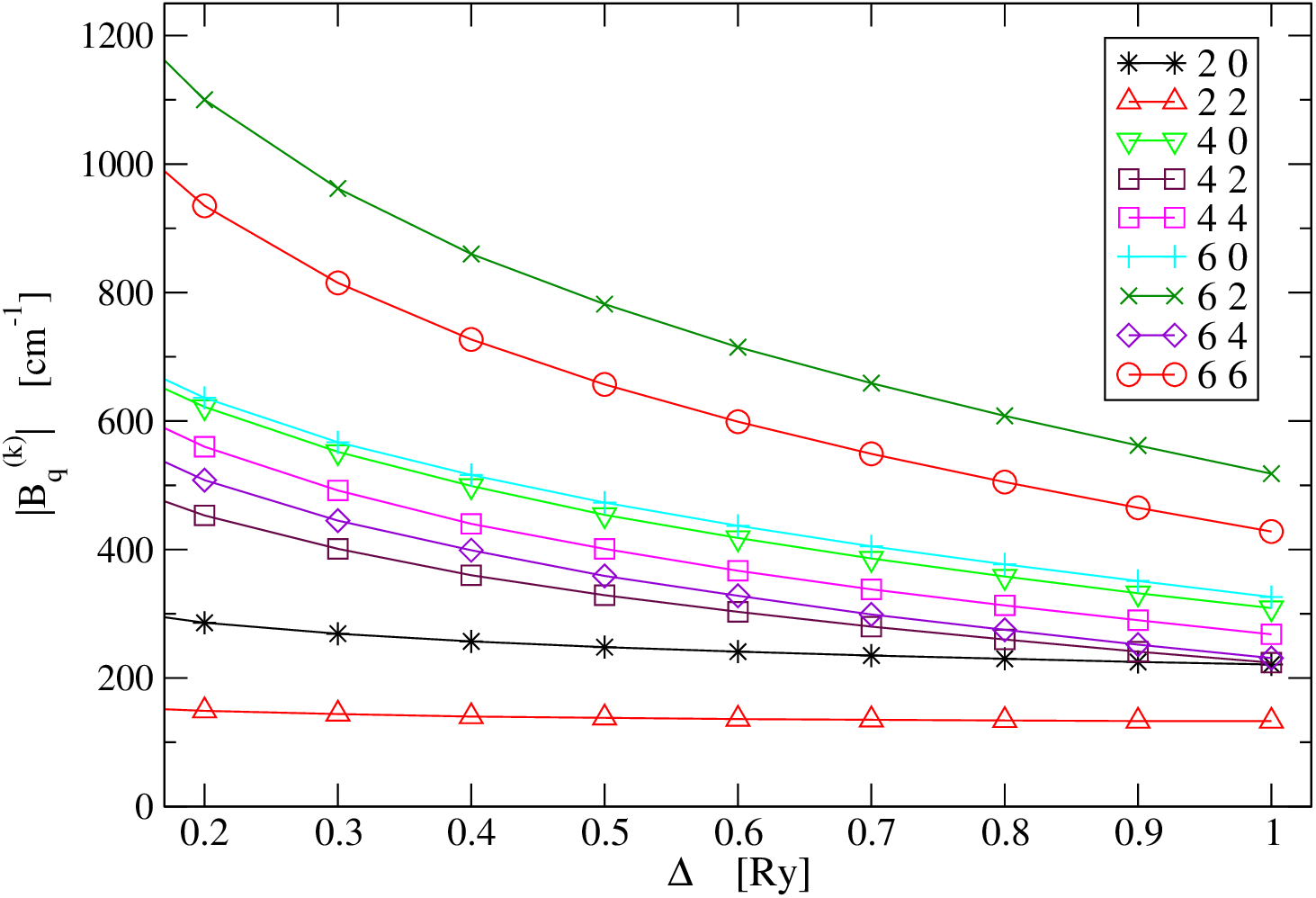}
\caption{(Color online) Nd:LaF$_3$. Dependence of the absolute value of the CFP  on the
hybridization parameter $\Delta$.}
\label{fig:babs_Nd}
\end{figure}

\begin{figure}
\includegraphics[width=12cm]{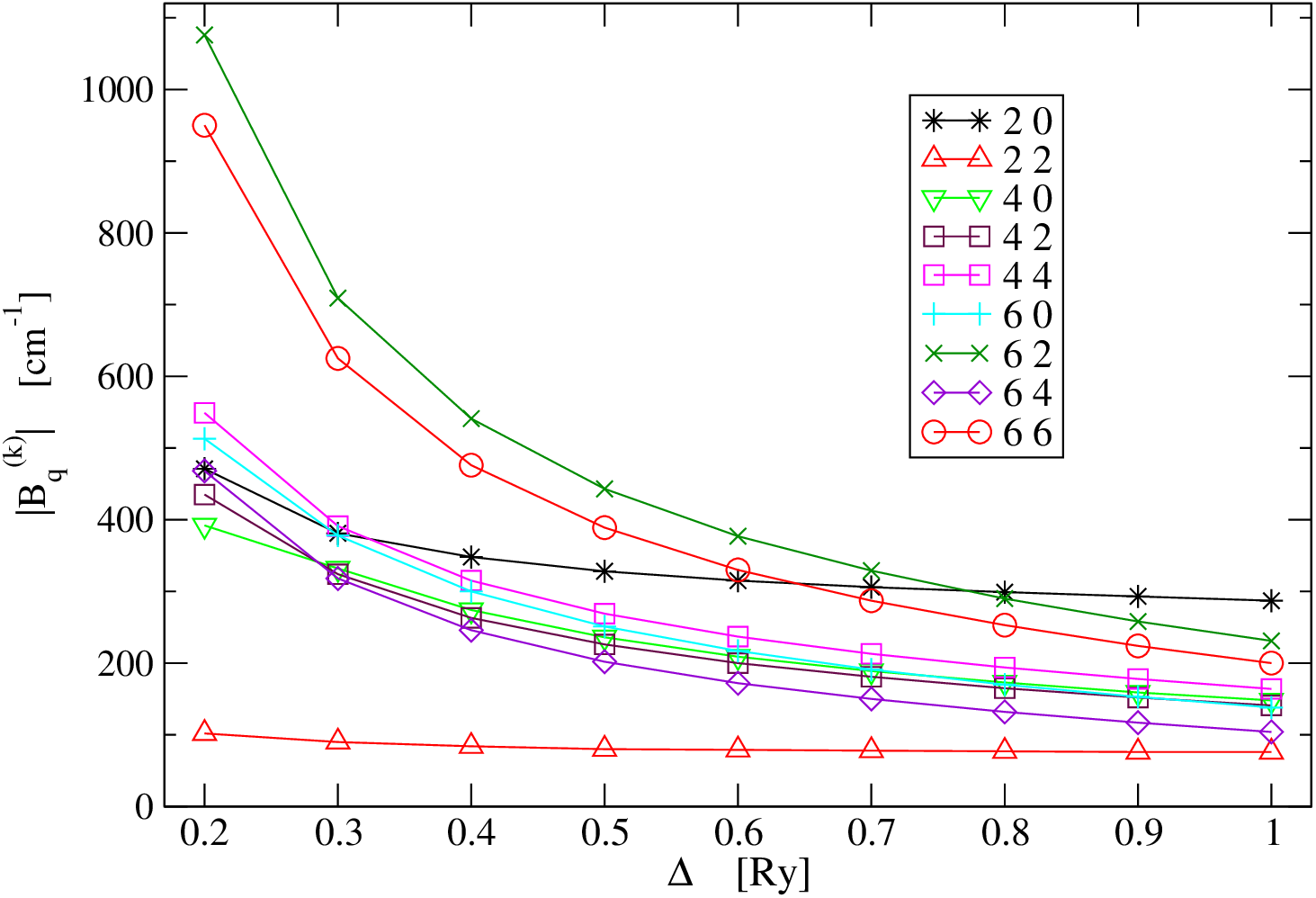}
\caption{(Color online) Er:LaF$_3$. Dependence of the absolute value of the CFP  on the
hybridization parameter $\Delta$. }
\label{fig:babs_Er}
\end{figure}

Tables \ref{tab:cfpr}, \ref{tab:cfpi} collect the CFP calculated for $\Delta$ = 0.4 Ry. They refer to
the orthogonal system depicted in Fig. \ref{fig:structure} with axis $z$ parallel to the hexagonal axis $c$, 
$x \parallel a$ and $y \perp a, c$. 
The dependence of CFP on the hybridization parameter is for $\Delta \geq 0.2$ smooth. For $\Delta < $ 0.2  
 the Wannier functions loose their atomic character and the CFP values become scattered. For different R the
character of $B^{(k)}_q(\Delta)$ dependence is similar, which is documented in Figs. \ref{fig:babs_Nd} and 
\ref{fig:babs_Er} for R = Nd and R = Er, respectively. In these figures the absolute values 
$|B^{(k)}_q|$ are shown. 

\begin{figure}
\includegraphics[width=12cm]{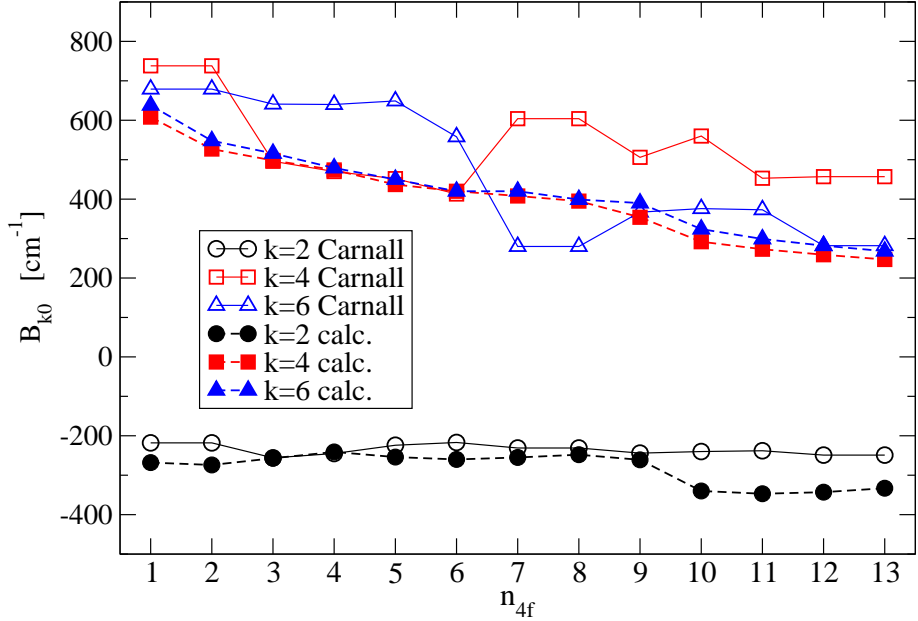}
\caption{(Color online) Parameters $B_{20}, B_{40},  B_{60}$ as function of the
number of $4f$ electrons. Full symbols correspond to results calculated with hybridization parameter $\Delta$ = 0.4 Ry.
Open symbols were determined using the CFP given by Carnall {\it et al.} \cite{carnall1}. }
\label{fig:Bk0}
\end{figure}

\begin{figure}
\includegraphics[width=12cm]{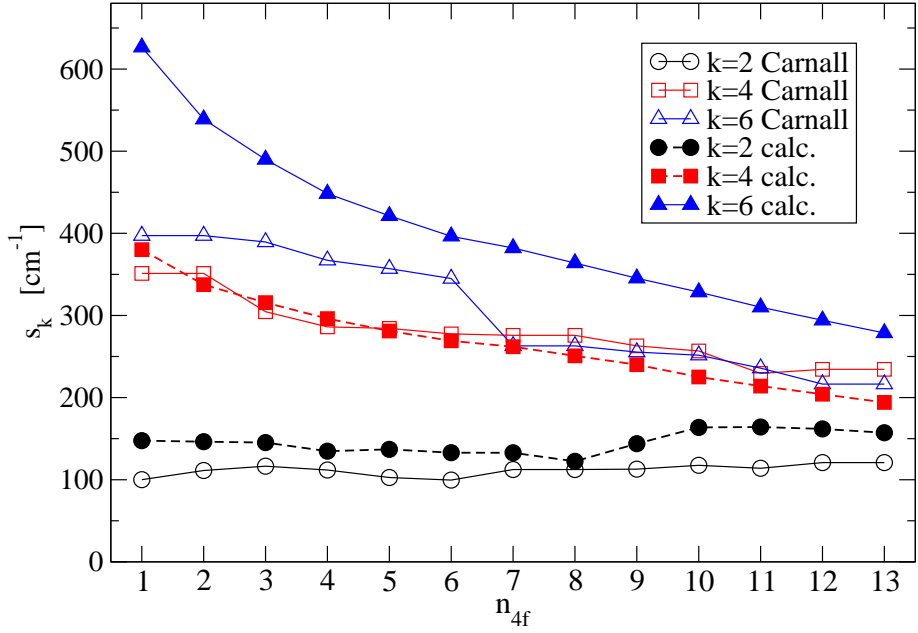}
\caption{(Color online) Quantity $s_k$ (see eq. \ref{eq:Sk}) as function of the
number of $4f$ electrons. Full symbols correspond to results calculated with hybridization parameter $\Delta$ = 0.4 Ry.
Open symbols are the CFP as given by Carnall {\it et al.} \cite{carnall1}. }
\label{fig:Sk}
\end{figure}

It is difficult to compare the calculated CFP with those obtained by Carnall {\it et al.} \cite{carnall1} by
fitting the optical data. Carnall's {\it et al.} CFP are real because of an assumed, approximate site symmetry. 
We can compare, however, the values of  $B_{q}^{(0)}$, which are real and, in addition, the 
quantities $s_k$
\begin{equation}
\label{eq:Sk}
   s_k = \left[\frac{1}{2k+1}\sum_{q=-k}^k |B_{q}^{(k)}|^2 \right]^{1/2},
\end{equation}
which were introduced by Leavitt \cite{leavitt}. The $q$-averages $s_k$ are real and invariant 
with respect to the rotation of the coordinate system.
This comparison is displayed in Figs. \ref{fig:Bk0} and \ref{fig:Sk}, respectively. 
   
\begin{table}
\caption{ Real part of the CFP in R:LaF$_3$  (units of cm$^{-1}$).
Hybridization parameter $\Delta$ = 0.4 Ry.}
\centering
\begin{tabular}{cccccccccccccc}
\hline
\hline
kq  & Ce& Pr &Pm  &  Nd  &  Sm  &  Eu  &   Gd &  Tb  &  Dy & Ho & Er & Tm & Yb \\
 \hline
20 &-268 &-274 &-257 &-241 &-254 &-260 &-255 &-248 &-261 &-340 &-347 &-343 &-333 \\
 22 & -52 & -61 &-104 &-102 &-106 & -90 &-101 & -79 &-130 & -90 & -79 & -76 & -76 \\
 40 & 608 & 527 & 498 & 474 & 437 & 420 & 408 & 395 & 354 & 292 & 273 & 259 & 247 \\
 42 &-344 &-306 &-285 &-271 &-259 &-247 &-240 &-229 &-222 &-214 &-204 &-195 &-185 \\
 44 &37 &28 &25 &10 & 0 &-6 & -12 & -19 &-9 & -16 & -16 & -15 & -14 \\   
 60 & 638 & 548 & 516 & 479 & 450 & 420 & 420 & 399 & 390 & 324 & 299 & 282 & 268 \\
 62 & 467 & 408 & 378 & 350 & 333 & 312 & 304 & 286 & 287 & 270 & 255 & 243 & 231 \\
 64 &-340 &-291 &-267 &-236 &-222 &-207 &-197 &-189 &-181 &-176 &-165 &-156 &-148 \\
 66 &-640 &-550 &-494 &-452 &-421 &-398 &-384 &-363 &-335 &-325 &-308 &-293 &-277 \\
\hline
 \hline
\end{tabular}
\label{tab:cfpr}
\end{table}

\begin{table}
\caption{ Imaginary part of the CFP in R:LaF$_3$.
Hybridization parameter $\Delta$ = 0.4 Ry.}
\centering
\begin{tabular}{cccccccccccccc}
\hline
\hline
kq  & Ce& Pr &Pm  &  Nd  &  Sm  &  Eu  &   Gd &  Tb  &  Dy & Ho & Er & Tm & Yb \\
\hline
 22 & 125 & 109 &93 &75 &55 &44 &33 &15 &23 &24 &25 &24 &21 \\   
 42 &-265 &-240 &-220 &-189 &-182 &-168 &-165 &-153 &-172 &-172 &-165 &-158 &-150 \\
 44 & 524 & 470 & 439 & 415 & 398 & 384 & 373 & 358 & 342 & 330 & 314 & 300 & 285 \\
 62 &1017 & 864 & 771 & 700 & 655 & 612 & 583 & 555 & 520 & 504 & 477 & 451 & 427 \\
 64 &-394 &-336 &-295 &-266 &-246 &-231 &-218 &-207 &-190 &-190 &-182 &-172 &-163 \\
 66 & 641 & 567 & 533 & 495 & 472 & 451 & 439 & 423 & 404 & 384 & 362 & 344 & 326 \\
 \hline
\hline
  \hline
\end{tabular}
\label{tab:cfpi}
\end{table}

\section{Discussion}
\label{sec:discussion}
We first discuss the DV\_ME method \cite{ogasawara} and its application to Pr:LaF$_3$ and Eu:LaF$_3$ as described in \cite{ishii,brik}.
The method calculates the multiplet splitting in two steps. In the first step the density functional theory based 
calculation is performed for a cluster consisting of the R$^{3+}$ ion and its ligands. The rest of the crystal
is taken into account by adding an effective Madelung potential to the Hamiltonian. In the second step the many-electron problem
is solved by selecting  as a basis  the cluster one-electron eigenstates with the dominant $4f$ character.  
Choice of the ligands to be included in the cluster and the cluster embedding are difficult problems (cf. Ref. \cite{ogasawara2} and Fig. 1 in \cite{ishii}). Another
problem is connected with the one-electron part of the many-electron Hamiltonian. The on-site interaction of the $4f$ electrons 
with the potential they create themselves is not accounted for.
This spurious self-interaction is inherent to most of the DFT based methods and it may 
seriously deform the calculated crystal field.
Also the one-electron part, taken from the discrete variational calculation already contains the electron correlation, which is for
the second time included in the many-electron Hamiltonian.
This are likely to be the reasons why the DV-ME calculated crystal field for Eu:LaF$_3$ is in poor agreement with the experiment
(cf. Table 1 in \cite{brik}). The agreement is significantly improved only when two semiempirical
parameters are introduced. In the present method the self-interaction is avoided by calculating the potential with the $4f$ states
in the core, where they contribute to the spherical part of the potential only (cf. 1st step in section \ref{sec:methods}) and
there is no double counting of the correlation.
 
The present results show convincingly that the method of CFP calculation, described in section \ref{sec:methods}, 
can be successfully used also if the R site has a low symmetry. The mean square deviation $\chi$ shown in Fig. \ref{fig:chi},
calculated for optimal value of the hybridization parameter is, with exception of Yb, smaller than 30 cm$^{-1}$. We 
point out that no attempt was done to regroup the experimental results, which would make $\chi$ even smaller. The hybridization
of the $4f$ states with the valence states of the fluorine plays an important role - as documented in Figs. 
\ref{fig:babs_Nd} and \ref{fig:babs_Er} increase of the hybridization (i.e. reduction of the parameter $\Delta$) changes
some of the CFP by as much as 100 \%. We note that large value of $\Delta$ corresponds to the
vanishing hybridization. The dependence of CFP on $\Delta$ is smooth, allowing thus an interpolation if there is a need.

With increasing number of the $4f$ electrons  the strength of crystal field the CFP of 4th and 6th order decrease,
while 2nd order CFP remain almost unchanged
(cf. Figs. \ref{fig:Bk0}, \ref{fig:Sk}), a similar behavior  we found for R$^{3+}$ impurities in the YAlO$_3$ \cite{novak1}.
The values of $s_k$ (see eq. \ref{eq:Sk}) and the $B_{0}^{(k)}$ compare well with those deduced from
the data of Carnall {\it et al.} \cite{carnall1} 
(Figs. \ref{fig:Bk0}, \ref{fig:Sk}).
The remaining difference may be connected with too high (16.7 \%) content of R atoms in the unit cell. To decrease this content would
require a supercell with several hundreds atoms, which is beyond our present computational possibilities. 
Other possible sources of the difference are the replacement of the actual $C_2$ local symmetry by $C_{2v}$
in the  Carnall {\it et al.} \cite{carnall1} analysis, uncertainty in experimental data (especially in cases of undetected
levels) and keeping $\Delta$ fixed at 0.4 Ry. 

 Previously we treated $\Delta$ as an adjustable parameter \cite{novak1,novak2,novak3}. 
In the present work we have also attempted to calculate $\Delta$ in a straightforward way described in section \ref{sec:delta}. 
The calculated $\Delta$ and $\Delta$ obtained from the fitting of experiment are in a fair agreement (Fig. \ref{fig:delta}),
indicating thus that the model can be made fully {\it ab-initio}.

\section{Conclusions}
The method to calculate crystal field parameters was successfully applied to
rare earth impurities in LaF$_3$, showing its applicability to R on sites with an arbitrary local
symmetry. Despite the ionic character of the compound the R($4f$) - F($2p,\;2s$) hybridization plays a
significant role. The parameter $\Delta$, which is needed to describe this hybridization, can be
calculated independently and the calculated value is in a fair agreement with value which fit
the experiment. The method thus has potential to be fully {\it ab-initio} i.e. only the atomic
composition and the crystal structure are needed to determine the crystal field. This would be
particularly useful in systems in which the optical absorption data are scarce or missing and 
in systems with defects and/or several different R ions.  The method has only moderate demands on the
computing facility and it is now freely accessible to public \cite{cfpwien}.


\end{document}